**Orbital-Driven Electronic Structure Changes and the Resulting Optical Anisotropy of the Quasi-Two-Dimensional Spin Gap Compound La$_4$Ru$_2$O$_{10}$**

S.J. Moon,[1] W.S. Choi,[1] S.J. Kim,[1] Y.S. Lee,[2] P.G. Khalifah,[3,4] D. Mandrus,[4] and T.W. Noh[1,*]

[1]*ReCOE & FPRD, Department of Physics and Astronomy, Seoul National University, Seoul 151-747, Korea*

[2]*Department of Physics, Soongsil University, Seoul 156-743, Korea*

[3]*Department of Chemistry, University of Massachusetts, Amherst, Massachusetts 01003, USA*

[4]*Materials Science and Engineering Division, Oak Ridge National Laboratory, Oak Ridge, Tennessee 37831, USA*

We investigated the electronic response of the quasi-two-dimensional spin gap compound La$_4$Ru$_2$O$_{10}$ using optical spectroscopy. We observed drastic changes in the optical spectra as the temperature decreased, resulting in anisotropy in the electronic structure of the spin-singlet ground state. Using the orbital-dependent hopping analysis, we found that orbital ordering plays a crucial role in forming the spin gap state in the non-one-dimensional material.

[*]E-mail address: twnoh@snu.ac.kr

In a one-dimensional (1D) periodic system, spin dimerization can occur in association with lattice distortion, leading to a stronger magnetic interaction within the dimers and opening a spin gap (spin-singlet ground state). This Peierls-type mechanism does not usually work in most materials with higher dimensions. Rather surprisingly, some three-dimensional (3D) spinels, such as $CuIr_2S_4$ and $MgTi_2O_4$, recently showed lattice dimerization accompanied by the spin gap opening. It was also found that the crystal structures of the spin-singlet states for $CuIr_2S_4$ and $MgTi_2O_4$ have octamer and helical patterns, respectively [1, 2]. In order to explain these spin-dimer containing complex superstructures, Khomskii and Mizokawa proposed a crucial role of orbital ordering in the formation of the Peierls states in these 3D spinels [3].

How can the orbital orderings drive the Peierls state in non-1D materials? Compared to other degrees of freedom in a solid, the orbitals have a unique feature. The orbitals have anisotropic shapes, so they can be oriented in particular directions. For degenerate orbitals in particular, each one points in a different direction, and their relative spatial orientation will determine the rate of electron hopping from an orbital of one ion to another orbital of a neighboring ion. With the appropriate orbital orderings, the relative orientations of the orbitals can be aligned in such a way to provide a higher hopping rate along quasi-1D chain directions, and lattice will be distorted accordingly to form superstructures of the dimers in the Peierls state [3, 4].

Up to this point, there is little experimental evidence of an anisotropic electronic response in non-1D spin gap materials. Most of the experimental investigations of $CuIr_2S_4$ and $MgTi_2O_4$ have been based on structural measurements using x-ray or neutron diffraction [1, 2]. Given the rather complex orbital superstructures in $CuIr_2S_4$ and $MgTi_2O_4$, it is very difficult to directly measure their quasi-1D electronic structures. In addition, the underlying superstructures are not aligned with the crystallographic axes, further hindering directional dependent studies.

Recently, La$_4$Ru$_2$O$_{10}$ was reported to have a spin gap despite its quasi-2D structure [5]. This unexpected magnetic state appeared with a first-order structural transition at $T_c$=160 K. Figures 1(a) and 1(b) show the crystal structure of La$_4$Ru$_2$O$_{10}$. The quasi-2D Ru-O network in the *bc*-plane consists of zigzag and linear chains of corner-sharing RuO$_6$ octahedra along the *b*- and *c*-axes, respectively. In the high-temperature (HT) phase above $T_c$, the bond distances between the neighboring octahedra are the same. However, in the low-temperature (LT) phase below $T_c$, the bond distances in the zigzag chain alternate. In Figs. 1(c)-(e), the arrows indicate the shortening of the bond distances in the *y*-direction and their lengthening in the *x*-direction.

Note that the Ru$^{4+}$ ions have four electrons in the triply degenerate $t_{2g}$ orbitals. Figures 1(c)-(e) depict the arrangements of $d_{xy}$, $d_{yz}$, and $d_{zx}$ orbitals. The electron hopping between them depends on the spatial orientation of the involved $t_{2g}$ orbitals. For example, while the hopping between the $d_{xy}$ orbitals occurs in the *x*- and *y*-directions, that between the $d_{yz}$ orbitals occurs in the *y*- and *z*-directions. Due to this orbital-dependent hopping, the local lattice distortion could accompany an orbital ordering and result in anisotropy in the electronic response. Further, unidirectional modulation along the *b*-axis could induce a global alignment of the macroscopic superstructure along the crystallographic axes. Therefore, La$_4$Ru$_2$O$_{10}$ could be an ideal model system to investigate the role of the orbitals in non-1D spin gap systems.

According to the Fermi's golden rule [6], the spectral weight (*SW*) of an optical transition will depend on the initial and final electronic density of states. Optical spectroscopy is a powerful experimental technique for investigating the electronic structures of strongly correlated electron materials [7, 8]. In this Letter, we investigated the temperature (*T*)- and polarization-dependent optical conductivity spectra $\sigma(\omega)$ of La$_4$Ru$_2$O$_{10}$. We found that significant *SW* redistribution occurs in the optical spectra of the quasi-2D *bc*-plane. Our *SW* analysis confirms that the $d_{zx}$ orbitals should be almost doubly occupied below $T_c$. Further, in the orbital-ordered LT phase, the *SW* analysis provides experimental evidence of how orbital ordering will result in strong electronic anisotropy, as expected for the singlet ground state.

La$_4$Ru$_2$O$_{10}$ single crystals were grown using the floating zone method [5]. Although the samples had two kinds of twin domains, they are related to each other by a two-fold rotation in the *bc*-plane. Therefore, the twin structure should not affect our polarization-dependent optical measurements on this plane. Optical microscopy studies confirmed that the *bc*-plane of our sample was very clean. Near normal incident reflectivity spectra $R(\omega)$ of the *b*- and *c*-axes were measured between 5 meV and 20 eV using polarized light [9]. The corresponding $\sigma(\omega)$ were obtained using Kramers-Kronig analysis.

Figure 2(a) shows the optical conductivity spectra along the *b*-axis, $\sigma_b(\omega)$. At room *T*, $\sigma_b(\omega)$ exhibits three optical transitions around 1.2, 2.1, and 3.4 eV. Since the strong peak at ~3.4 eV can be assigned as a charge transfer excitation from the O 2*p* to Ru $t_{2g}$ bands [10], the peaks below 2.5 eV should be associated with the interatomic transitions between the Ru 4*d* $t_{2g}$ orbitals. At the lowest *T*=10 K, the lower energy peak $\alpha_b$ is suppressed, while the higher energy peak $\beta_b$ is enhanced, implying a change in electronic structure. The *SW* redistribution occurs most significantly across $T_c$ (see Fig. 3.), where the phonon spectra also exhibit drastic changes as shown in the inset of Fig. 2(a). These results indicate that the change in electronic structure is closely related to the lattice distortion and concomitant spin gap opening.

In order to elucidate the changes of electronic structure, we should know the correct orbital and spin states in the LT phase of La$_4$Ru$_2$O$_{10}$. Note that there have been discussions on the nature of the spin gap. For $T > T_c$, the Ru$^{4+}$ ion should have the usual $t_{2g\uparrow}^3 t_{2g\downarrow}^1$ low-spin state with *S*=1. Khalifah *et al.* initially suggested that the LT phase might be in a $t_{2g\uparrow}^2 t_{2g\downarrow}^2$ spin state with *S*=0, caused by a sufficiently strong crystal field splitting due to the lattice distortion [5]. A schematic diagram is shown in Fig. 2(b). Later, based on the results of x-ray absorption spectroscopy and a first-principles calculation, Wu *et al.* reported that the $Ru^{4+}$ ions could remain in the $S=1$ spin state with $t_{2g\uparrow}^3 t_{2g\downarrow}^1$ in the LT phase, as shown in Fig. 2(c) [11]. They attributed the spin gap state to the formation of local spin-singlet Ru-Ru dimers, which comes from a strongly anisotropic exchange interaction.

To clarify this issue, we performed the *SW* analysis on the basis of an orbital correlation. In the HT phase, each *d* orbital is partially filled, so all optical transitions between the same orbitals of neighboring Ru ions are allowed. Conversely, in the LT phase, the difference in orbital occupation will result in *SW* changes. As mentioned previously, the lattice distortion of the LT phase involves the alternation of the short *y*- and long *x*-directional bonds in zigzag chains along the *b*-axis. Due to the associated crystal field splitting, the $d_{zx}$ orbital is lower in energy than the $d_{xy/yz}$ orbitals. As shown in Fig. 2(c), the $d_{zx}$ orbital then becomes doubly occupied and the other two orbitals are singly occupied. In this LT phase, the optical transition between the doubly occupied $d_{zx}$ orbitals becomes forbidden, and those between the $d_{xy/yz}$ orbitals should be enhanced [12]. The expected *SW* changes are well reproduced in Fig. 2(a): a drastic suppression of the peak $\alpha_b$ and an enhancement of the peak $\beta_b$ below $T_c$. *Based on this analysis, we assign the peak $\alpha_b$ as the transition between the $d_{zx}$ orbitals and the peak $\beta_b$ as the transitions between the $d_{xy/yz}$ orbitals*. These assignments are consistent with the predictions of the first-principles calculation [11]. Therefore, in the LT phase, the $d_{zx}$ orbital should be almost doubly occupied and the associated local spin state of $Ru^{4+}$ ion remains as the $S=1$.

Now, let us investigate how the relative orbital orientations in the orbital ordered state resulted in anisotropic changes of the *SW*. In Fig. 1(a), the zigzag chains along the *b*-axis point in either the *x*- or the *y*-direction. In order to observe the anisotropic nature of electron structure, it would be the most appropriate to perform polarization-dependent optical measurements on the *ab*-plane. However, due to the layered structure of $La_4Ru_2O_{10}$, it is difficult to prepare samples with a sufficiently good *ab*-plane. Furthermore, twinning at low *T* makes it impossible to individually measure the optical responses in the *x*- and *y*-directions, as they are superimposed by the twin operator. Therefore, the optical measurements had to be performed on the *bc*-plane.

Figures 3(a) and 3(b) show the *T*-dependent $\sigma_b(\omega)$ and $\sigma_c(\omega)$, respectively. Qualitative changes in both spectra are quite similar: the *SW* of the peak $\alpha$ ($\beta$)becomes suppressed (enhanced) with decreasing *T*. However, there are significant differences in the amount of the *SW* changes along the *b*- and *c*-axes. At room *T*, $\sigma_b(\omega)$ and $\sigma_c(\omega)$ are nearly identical, indicating nearly isotropic electron hopping between the *d* orbitals in the *bc*-plane. Below $T_c$, the enhancement of the higher energy peak $\beta$ in $\sigma_b(\omega)$ is much larger than that in $\sigma_c(\omega)$, suggesting the anisotropy in the electronic structure.

To obtain further insight, we estimated the total polarization-dependent *SW* of the interatomic *d-d* transitions, $SW^{\alpha+\beta}$, i.e., the sum of the *SW* of the peaks $\alpha$ and $\beta$. We subtracted the contributions of the charge transfer transitions and integrated the remaining $\sigma_b(\omega)$ and $\sigma_c(\omega)$ between 0 and 3.5 eV. We found that the $SW^{\alpha+\beta}$ along the *b*-axis became larger than that along the *c*-axis. Since the contribution of the *SW* of the peak $\alpha$ is almost suppressed at the lowest *T*, the changes in *SW* should be closely linked to those for the peak $\beta$. To obtain more quantitative information, we fit $\sigma(\omega)$ using the Lorentz oscillator model and estimated the *SW* of each peak: the result is shown in Fig. 3(c). Note that the decreased *SW* of the peaks $\alpha$ for both *b*- and *c*-axes transferred to the peak $\beta$ mostly for the *b*-axis. The total amount of transferred *SW* from the *c*- to the *b*-axis should be related to the change of the occupation number in the orbital-ordered state. As shown in the inset of Fig. 3(c), at the lowest *T*, the *SW* of the peak $\beta$ for the *b*-axis, $SW_b^\beta$, became about 1.9 times larger than $SW_c^\beta$.

To account for the observed anisotropic *SW* quantitatively, we considered a simple orbital-dependent hopping model. From the Fermi's golden rule, the optical *SW* of an

interatomic *d-d* transition can be written as

$$SW \propto \left| \sum_m \langle \psi_f | p_m \rangle \langle p_m | \psi_i \rangle \right|^2 / \Delta, \qquad (1)$$

by assuming that the intermediate energy $\Delta$ remains almost unchanged [13]. The initial state $\psi_i$ and the final state $\psi_f$ correspond to the occupied $t_{2g}$ orbitals at one Ru site and the unoccupied $t_{2g}$ orbitals at the neighboring Ru site, respectively. And $|p_m\rangle$ ($m=x, y, z$) represents the O 2*p* orbitals. As the O 2*p* orbitals are nearly fully occupied, the terms in Eq. (1) should be dependent on the relative orientation of the Ru $t_{2g}$ orbitals. In the HT phase, each of the three $t_{2g}$ orbitals contributes equally to the hopping process. Conversely, in the LT phase, only the $d_{xy}$ and $d_{yz}$ orbitals contribute to the hopping process. This occupation change due to orbital ordering can increase the hopping rate of the $d_{xy}$ and $d_{yz}$ orbitals by 1.5 times. In addition, the hopping rate also depends on the bond length and angle [14]:

$$\left| \sum_m \langle \psi_f | p_m \rangle \langle p_m | \psi_i \rangle \right|^2 \propto \left| \frac{\cos\theta}{d^{3.5}} \right|^2, \qquad (2)$$

where $\theta$ and $d$ are the Ru-O-Ru bond angle and the Ru-O bond length, respectively.

Initially, we estimated the values of *SW* in the *x*-, *y*-, and *z*-directions using Eqs. (1) and (2). For simplicity, we assumed that the values of $\left| \sum_m \langle \psi_f | p_m \rangle \langle p_m | \psi_i \rangle \right|^2$ for the HT phase were the same along the *x*-, *y*-, and *z*-directions. Using literature values for $\theta$ and $d$ [15], we estimated the *SW* of the LT phase along the optically measured crystallographic *b*- and *c*-axes. Without orbital ordering, we found that the transfer of $SW^{\alpha+\beta}$ from the *c*-axis to the *b*-axis is only about 4%, and that $SW_b^\beta / SW_c^\beta$ remains almost the same as that of the HT phase. These changes in *SW* are too small to explain our experimental observations, seen in the insets of Figs.

3(b) and 3(c). On the other hand, if we include the orbital ordering, $SW^{\alpha+\beta}$ along the *b*-axis (*c*-axis) increases (decrease) by about 25%. These values are consistent with the experimentally observed changes of $SW^{\alpha+\beta}$, shown in the inset of Fig. 3(b). Further, this simple orbital ordering based calculation yielded that $SW_b^{\beta}$ should be about 1.8 times larger than $SW_c^{\beta}$, which agrees quite well with the experimental value of 1.9, shown in the inset of Fig. 3(c). These results indicate the crucial role of orbital ordering in generating the electronic anisotropy of $La_4Ru_2O_{10}$.

At this moment, we want to point out a couple of important points. First, note that the 1.9 times enhancement in $SW^{\beta}_b/SW^{\beta}_c$ indicates that the $SW^{\beta}$ along the *y*-axis should be about 3.3 times larger than that along the *x*-axis. This large value is consistent with the fact that the LT phase of $La_4Ru_2O_{10}$ should be in a dimerized state. Since the magnetic properties studies demonstrate that the LT phase is a non-magnetic state [5], it should be a spin-singlet dimer state. Second, it is worthwhile to mention that the orbital/spin ordering correlation persists above $T_c$. As displayed in the inset of Fig. 3(c), the in-plane anisotropy develops rather gradually. This contrasts the abrupt structural change at $T_c$, suggested by the phonon mode changes in the inset of Fig. 2(a). This indicates that short-range orbital/spin fluctuations still remain at a rather wide *T* region above $T_c$. Further studies on this orbital /spin correlation in the HT phase are highly desirable.

In summary, we investigated the temperature- and polarization-dependent optical conductivity spectra of $La_4Ru_2O_{10}$. We observed robust spectral changes with temperature, which were associated with the formation of a spin-singlet state accompanied by a lattice distortion. In the singlet ground state, the optical response was found to be strongly anisotropic, effectively resulting in a one-dimensional-like response. Based on the orbital-dependent hopping analysis, we showed that the orbital ordering plays a crucial role in the formation of the spin-singlet state.

We acknowledge valuable discussions with D.I. Khomskii, J.D. Denlinger, and J. Yu. This work was financially supported by Creative Research Initiative Program (Functionally Integrated Oxide Heterostructure) of MOST/KOSEF. YSL was supported by the Soongsil University Research Fund. Experiments at PLS were supported in part by MOST and POSTECH. Research at ORNL was sponsored by the Division of Materials Science and Engineering, Office of Basic Energy Sciences, U.S. DOE.

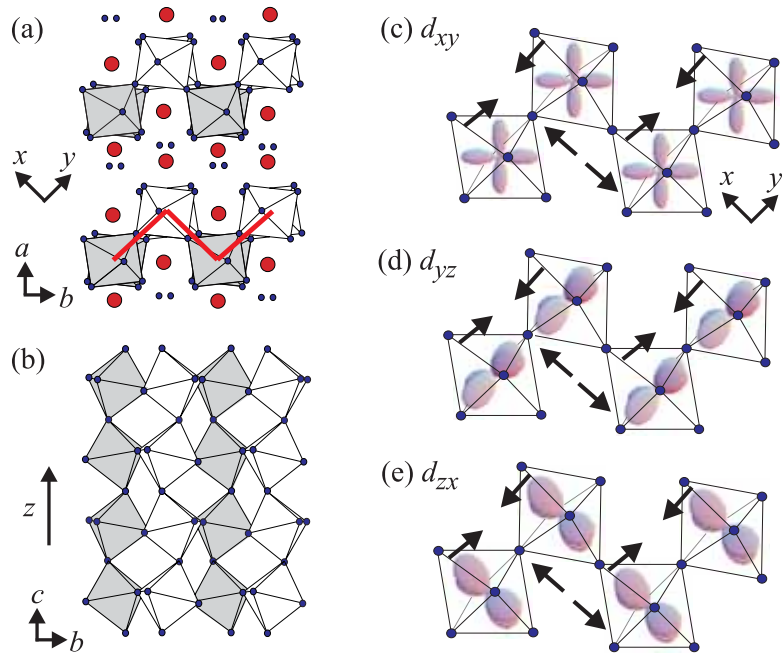

Fig. 1 (color online). The crystal structure of $La_4Ru_2O_{10}$ in the crystallographic (a) *ab*-plane and (b) *bc*-plane. The large red and small blue dots denote the La and O ions, respectively. The shaded $RuO_6$ octahedra are at the lower positions on the *a*-axis. Note that the zigzag chains (along the red line) are formed along the *b*-axis. (c), (d), and (e) display the Ru $d_{xy}$, $d_{yz}$, and $d_{zx}$ orbitals on the lattice, respectively. The arrows represent the distortions of the Ru-O bonds due to the structural transition. The local lattice distortion results in lengthening and shortening of the Ru-Ru bonds along the zigzag chains, resulting in lattice dimerization. The *x*- (*y*-) direction is chosen parallel to the direction of the long (short) Ru-Ru bonds.

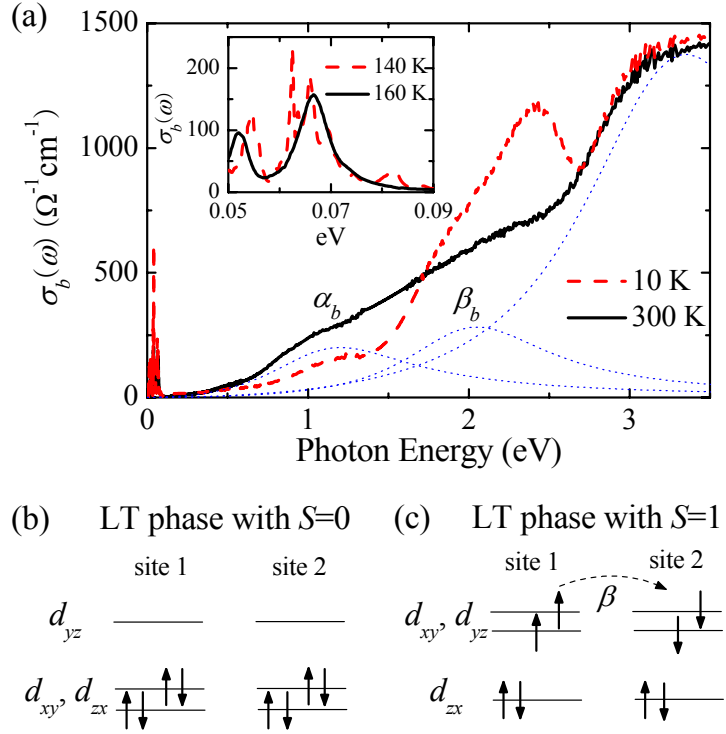

Fig. 2 (color online). (a) $T$-dependent optical conductivity spectra $\sigma_b(\omega)$ along the $b$-direction. The interatomic $d$-$d$ transitions are labeled as $\alpha_b$ and $\beta_b$. The blue dotted lines represent the contribution of the peaks $\alpha_b$, $\beta_b$, and the charge transfer excitation peak in the Lorentz oscillator fit. The inset shows the phonon spectra between 0.05 and 0.09 eV across the structural transition temperature. (b) and (c) show the spin/orbital configurations of the neighboring $t_{2g}$ orbitals in the suggested LT phases with $S=0$ (Ref. 5) and $S=1$ (Ref. 11), respectively. In the $S=0$ LT phase, no optical transition is allowed. By contrast, in the $S=1$ LT phase, the optical transition between the $d_{xy/yz}$ orbitals (i.e., peak $\beta$) is allowed, as indicated by the dashed arrow.

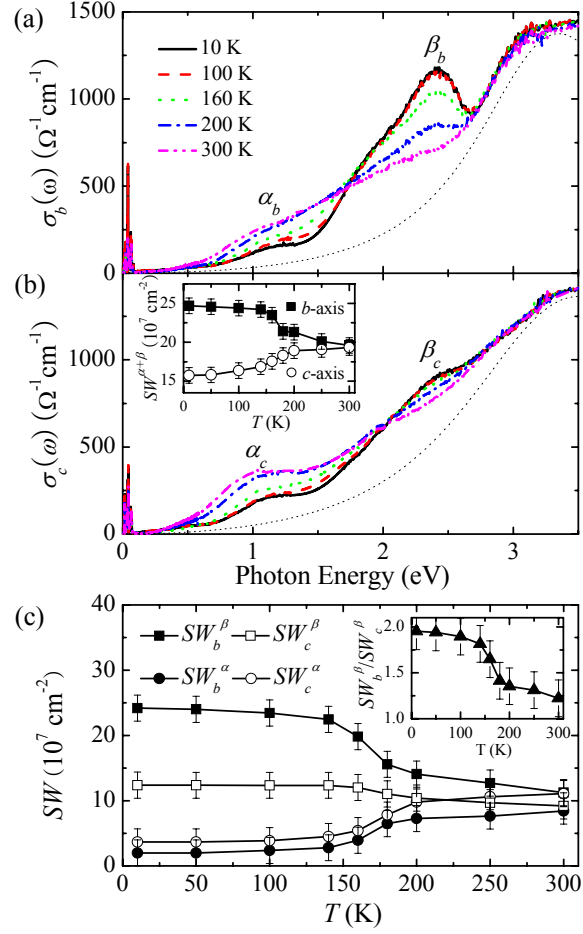

Fig. 3 (color online). *T*-dependent (a) $\sigma_b(\omega)$ and (b) $\sigma_c(\omega)$. The interatomic *d-d* transition between the neighboring $d_{zx}$ ($d_{xy/yz}$) orbitals is labeled as the peak $\alpha$ ($\beta$). The thin dotted lines represent the contributions from the charge transfer excitation. The inset in (b) shows the total spectral weights of the interatomic *d-d* transitions, i.e., the contribution from both of the peaks $\alpha$ and $\beta$, $SW^{\alpha+\beta}$. The solid squares and open circles indicate the *T*-dependent $SW^{\alpha+\beta}$ along the *b*- and *c*-axes, respectively. (c) *T*- and polarization-dependent changes of the *SW* of the peaks $\alpha$ and $\beta$, obtained from the Lorentz oscillator model fit. The inset shows the ratio of $SW_b^\beta$ to $SW_c^\beta$, which represents the optical anisotropy in peak $\beta$ in the *bc*-plane.